# Relations for Massive Spinors

Clemens Heuson

*Ulrichstr. 1, D-87493 Lauben, Germany*
e-mail: clemens.heuson@freenet.de

Recently introduced massive spinors are written as 2-vectors consisting of two massless spinors with opposite helicities. With this notation a couple of relations between them can be derived easily, entirely avoiding the spinor indices. The high energy limit of three point amplitudes is discussed shortly. Finally we add some comments on recursion relations with massive particles.

## 1. Introduction

The spinor helicity formalism, see for example the reviews in [1],[2],[3],[4], has boosted the calculation of amplitudes in particle physics. Amplitudes that could not be done even with computers can now be calculated with much less effort. But the advantage is not only on the side of faster calculations. Feynman diagrams, relying on manifest Lorentz invariance of the Lagrangian, describe massless spin one bosons like photons or gluons by a vector with four components and a massless spin two graviton by a symmetric rank two tensor with ten components. Massless states however have only two helicities, positive and negative. This redundancy in the description necessarily requires gauge and diffeomorphism invariance of the gauge and graviton field. A redundancy appears already at the level of scalar fields in the form of field redefinitions [4]. Graviton physics becomes very complicated with the Lagrangian formalism, see for example the complicated term for the interaction between fermions and gravitons in [5] or the infinitely many terms for graviton selfinteractions. Compare this with the simple expressions for gravity amplitudes in literature [1-4].

The spinor helicity formalism had one limitation, it was only valid for massless particles and thus could only serve as an approximation for massive particles in the high energy regime, where their mass can be neglected. Massive spinor helicity variables were first introduced by several authors, see for example [6] and related work. In their seminal work Arkani-Hamed, Huang and Huang extended the spinor helicity formalism to amplitudes for all masses and spins [7]. For massless particles with momentum $p^\mu = (P \; 0 \; 0 \; P)$ the little group is $U(1)$, while for massive particles with momentum $p^\mu = (m \; 0 \; 0 \; 0)$ in the restframe the little group is $SU(2)$. Massive particles are described as 2x2 matrices $\lambda_\alpha^I, \tilde{\lambda}_{\dot\alpha}^I$ with $\alpha, \dot\alpha$ denoting the $SL(2,\mathbb{C})$ indices and $I,J$ denoting the $SU(2)$ spin indices. Many following papers have investigated amplitudes within this new formalism, see for example [8],[9],[10],[11] and many others.

Here we make a minor step in formulating massive spinors as 2-vectors consisting of two massless spinors with opposite helicities. Of course this is already implicit in [7], [6] and was also suggested in [12]. We shall find that many relations between massive spinors can be derived easily with this. The high energy limit of three particle amplitudes is discussed. Finally some comments on recursion relations are made.

## 2. Relations between massive spinors by 2-vectors

We use the representation of massive spinors given in [10],[11] with mostly minus metric and the four momentum given by

$$p_\mu = (E \quad P\sin(\theta)\cos(\phi) \quad P\sin(\theta)\sin(\phi) \quad P\cos(\theta)) \qquad (1)$$

Using the Pauli matrices, the momentum can be written in spinor notation $p = p_{\alpha\dot\alpha} = p_\mu \sigma^\mu$ and $\bar{p} = p^{\dot\alpha\alpha} = p_\mu \bar\sigma^\mu$,

$$p = p_{\alpha\dot\alpha} = \begin{pmatrix} E + P(cc - ss^*) & 2Pcs^* \\ 2Pcs & E - P(cc - ss^*) \end{pmatrix}, \quad \bar{p} = p^{\dot\alpha\alpha} = \begin{pmatrix} E - P(cc - ss^*) & -2Pcs^* \\ -2Pcs & E + P(cc - ss^*) \end{pmatrix} \qquad (2)$$

where as usual $c = \cos(\theta/2)$, $s = \sin(\theta/2)e^{i\phi}$, $s^* = \sin(\theta/2)e^{-i\phi}$. Now we write the massive spinors given in [7],[10],[11] as 2-vectors for example $|i^I\rangle = |p_i^I\rangle = (|i\rangle \quad |n_i\rangle)$. The massless spinor $|i\rangle$ scales with $\sqrt{E_i + P_i}$ and is denoted in the same way as the corresponding spinor for massless particles scaling with $\sqrt{2E_i}$ because they are equal in the high energy limit. This should in general not generate any confusion, since one knows for any amplitude which particles are massive and which are massless. One could attach an index 0 for massless particles if necessary. The second massless spinor $|n_i\rangle$, (memo n = null spinor) was denoted as $|\eta_i\rangle$ in [7],[10],[11], and scales with $\sqrt{E_i - P_i}$ and therefore vanishes in the high energy limit. We now write down all possible massive spinors in the 2-vector notation,

$$|i^I\rangle = (|i\rangle \ \ |n_i\rangle) \ , \ \langle i^I| = (\langle i| \ \ \langle n_i|) \ , \ |i^I] = (-|n_i] \ \ |i]) \ , \ [i^I| = (-[n_i| \ \ [i|) \tag{3}$$
$$|i_I\rangle = (|n_i\rangle \ \ -|i\rangle) \ , \ \langle i_I| = (\langle n_i| \ \ -\langle i|) \ , \ |i_I] = (|i] \ \ |n_i]) \ , \ [i_I| = ([i| \ \ [n_i|)$$

where the spinors i and $n_i$ are explicitly given as

$$|i\rangle = \sqrt{E_i + P_i}\begin{pmatrix} c_i \\ s_i \end{pmatrix} \ , \ |n_i\rangle = \sqrt{E_i - P_i}\begin{pmatrix} -s_i^* \\ c_i \end{pmatrix} \ , \ \langle i| = \sqrt{E_i + P_i}\begin{pmatrix} s_i \\ -c_i \end{pmatrix} \ , \ \langle n_i| = \sqrt{E_i - P_i}\begin{pmatrix} c_i \\ s_i^* \end{pmatrix} \tag{4}$$
$$|i] = \sqrt{E_i + P_i}\begin{pmatrix} s_i^* \\ -c_i \end{pmatrix} \ , \ |n_i] = \sqrt{E_i - P_i}\begin{pmatrix} c_i \\ s_i \end{pmatrix} \ , \ [i| = \sqrt{E_i + P_i}\begin{pmatrix} c_i \\ s_i^* \end{pmatrix} \ , \ [n_i| = \sqrt{E_i - P_i}\begin{pmatrix} -s_i \\ c_i \end{pmatrix}$$

One doesn't need to write the explicit $SL(2,\mathbb{C})$ indices anymore, which simplifies many formulas. They can be reinserted easily by recalling that in angle brackets $\langle i\ j\rangle$ the index $\alpha$ is descending from left to right, while for square brackets $[i\ j]$ the index $\dot\alpha$ is ascending from left to right. In Lorentzinvariant amplitudes these indices are always contracted. From the explicit representation in (4) one can derive two important relations
(memo: negative/positive helicity spinors give a minus/plus sign).

$$\langle i\ n_i\rangle = -m_i \ , \ [i\ n_i] = +m_i \tag{5}$$

Therefore in rest of this paper we don't need the explicit representation given in (4) anymore. A further explicit representation was provided in [8], [9]. In appendix A still another representation with the standard momentum $p^\mu$ given by (1) is written down. The momentum in spinor form (2) can be written in the following form, as can be checked with the explicit spinors in (4)

$$p_i = |i^I\rangle[i_I| = -|i_I\rangle[i^I| = |i\rangle[i| + |n_i\rangle[n_i| \ , \ \bar{p}_i = |i_I]\langle i^I| = -|i^I]\langle i_I| = |i]\langle i| + |n_i]\langle n_i| \tag{6}$$

With the 2-vector notation we get using a dot product between the vectors:
$p_i = |i^I\rangle[i_I| = (|i\rangle \ \ |n_i\rangle)\cdot([i| \ \ [n_i|) = |i\rangle[i| + |n_i\rangle[n_i|$. The square of a momentum can be obtained using (5):
$p_i \cdot p_i = \frac{1}{2}\text{Tr}\{p_i \cdot \bar{p}_i\} = \frac{1}{2}\text{Tr}\{(|i\rangle[i| + |n_i\rangle[n_i|)(|i]\langle i| + |n_i]\langle n_i|)\} = \frac{1}{2}([i\ n_i]\langle n_i\ i\rangle + [n_i\ i]\langle i\ n_i\rangle) = m_i^2$. The action of momentum on a spinor now goes as: $p_i|i^I] = (|i\rangle[i| + |n_i\rangle[n_i|)(-|n_i] \ \ |i]) = -m_i(|i\rangle \ \ |n_i\rangle) = -m_i|i^I\rangle$. Square or angle brackets require a tensor product between 2-vectors, for example:

$$\langle i^J\ i^K\rangle = (\langle i| \ \ \langle n_i|)(|i\rangle \ \ |n_i\rangle) = \begin{pmatrix} \langle i\ i\rangle & \langle i\ n_i\rangle \\ \langle n_i\ i\rangle & \langle n_i\ n_i\rangle \end{pmatrix} = \begin{pmatrix} 0 & -m_i \\ m_i & 0 \end{pmatrix} = -m_i\,\epsilon^{JK}$$

In the same manner using (3) and (5), the following relations can be obtained:

$$\langle i^J\ i^K\rangle = -m_i\,\epsilon^{JK} \ , \ \langle i_J\ i_K\rangle = +m_i\,\epsilon_{JK} \ , \ [i^J\ i^K] = +m_i\,\epsilon^{JK} \ , \ [i_J\ i_K] = -m_i\,\epsilon_{JK} \tag{7}$$
$$\langle i_J\ i^K\rangle = +m_i\,\delta_J^K \ , \ \langle i^J\ i_K\rangle = -m_i\,\delta_K^J \ , \ [i_J\ i^K] = -m_i\,\delta_J^K \ , \ [i^J\ i_K] = +m_i\,\delta_K^J$$
$$\langle i^I\ i_I\rangle = -2m_i \ , \ [i^I\ i_I] = +2m_i \ , \ |i^I\rangle\langle i_I| = m_i\,\delta_\alpha^\beta \ , \ |i_I]\,[i^I| = m_i\,\delta_{\dot\beta}^{\dot\alpha}$$
$$2p_i \cdot p_j = \langle i^I|p_j|i_I] = = \langle j^I|p_i|j_I] = \langle i^I\ j^K\rangle[j_K\ i_I] \ , \ \langle j^K\ i\rangle\langle j_K\ i\rangle = [j^K\ i][j_K\ i] = 0$$
$$\langle j^K\ i^I\rangle\langle i_I\ j_K\rangle = [j^K\ i^I][i_I\ j_K] = -2m_i m_j \ , \ \langle i^I|p_j|i^J]\langle i_I|p_j|i_J] = \langle j^I|p_i|j^J]\langle j_I|p_i|j_J] = 2m_i^2 m_j^2$$
$$p_i|i^I] = -m_i|i^I\rangle \ , \ \bar{p}_i|i^I\rangle = -m_i|i^I] \ , \ \langle i^I|p_i = [i^I|m_i \ , \ [i^I|\bar{p}_i = \langle i^I|m_i$$

Most of these relations were of course described in literature [7-11], but derived here in a simple way using 2-vectors. We note some properties of the $\epsilon$ tensor and of $SU(2)$ vectors, which will be crucial later. The $\epsilon$ tensor is defined as $\epsilon^{JK} = -\epsilon_{JK} = \begin{pmatrix} 0 & 1 \\ -1 & 0 \end{pmatrix}$. Raising and lowering of indices as well as products between $SU(2)$ vectors goes as follows:

$$a_I = a^J \epsilon_{JI}, \quad a^I = a_J \epsilon^{JI}, \quad a^J b_J = -a_J b^J, \quad a^J a_J = 0 = a_J a^J \tag{8}$$

Note also, if $a^I = \begin{pmatrix} a^1 & a^2 \end{pmatrix}$ then $a_I = \begin{pmatrix} a^2 & -a^1 \end{pmatrix}$, if $a_I = \begin{pmatrix} a_1 & a_2 \end{pmatrix}$ then $a^I = \begin{pmatrix} -a_2 & a_1 \end{pmatrix}$.

Now consider the helicity operator defined as $h = \frac{\vec{p} \cdot \vec{\sigma}}{2|\vec{p}|} = \frac{1}{2}\begin{pmatrix} -(cc - ss^*) & -2cs^* \\ -2cs & (cc - ss^*) \end{pmatrix}$. Acting on the explicit spinors in (4) gives the result, that $|i\rangle, |i]$ have the same helicity as their massless counterparts, but the spinors $|n_i\rangle, |n_i]$ just have the opposite helicities. It can also be seen from the explicit form in (4) that for example $|n_i] \sim |i\rangle$ and therefore these spinors should have the same helicity. In summary we get:

$$h|i\rangle = -\frac{1}{2}|i\rangle, \quad h|i] = +\frac{1}{2}|i], \quad h|n_i\rangle = +\frac{1}{2}|n_i\rangle, \quad h|n_i] = -\frac{1}{2}|n_i].$$

## 3. Three particle amplitudes and high energy limit

In this section we consider three point vertices for particles with mass, the three legs are called i, j, k. Momentum conservation demands $p_i + p_j + p_k = 0$ or explicitly: $|i\rangle[i| + |n_i\rangle[n_i| + |j\rangle[j| + |n_j\rangle[n_j| + |k\rangle[k| + |n_k\rangle[n_k| = 0$. Multiplying from left with $\langle j|, \langle n_j|, \langle \xi|$ ($\xi$ = arbitrary spinor) and from right with $|k]$ gives the following equations, using that $n_i$ scales with $m_i$ and therefore can be neglected at first order in the high energy limit.

$$\langle j\,i\rangle[i\,k] \approx 0 + O(m^2) \tag{9}$$

$$\langle n_j\,i\rangle[i\,k] \approx -m_j[j\,k] + O(m^3) \tag{10}$$

$$\langle \xi\,j\rangle[j\,k] \approx -\langle \xi\,i\rangle[i\,k] + O(m^2) \tag{11}$$

Starting from $\bar{p}_i + \bar{p}_j + \bar{p}_k = 0$ or $|i]\langle i| + |n_i]\langle n_i| + |j]\langle j| + |n_j]\langle n_j| + |k]\langle k| + |n_k]\langle n_k| = 0$ similar relations can be obtained. Multiplying with $[j|, [n_j|, [\xi|$ from left and $|k\rangle$ from right one obtains using the scaling of $n_i$:

$$[j\,i]\langle i\,k\rangle \approx 0 + O(m^2) \tag{12}$$

$$[n_j\,i]\langle i\,k\rangle \approx m_j\langle j\,k\rangle + O(m^3) \tag{13}$$

$$[\xi\,j]\langle j\,k\rangle \approx -[\xi\,i]\langle i\,k\rangle + O(m^2) \tag{14}$$

The high energy limit of factors in (4) is $\sqrt{E+P} \approx \sqrt{2E}\left(1 - \frac{m^2}{8E^2}\right)$ and $\sqrt{E-P} = \frac{m}{\sqrt{E+P}} \approx \sqrt{2E}\frac{m}{2E}\left(1 + \frac{m^2}{8E^2}\right) \approx \frac{m}{\sqrt{2E}}$, where we expanded $P = \sqrt{E^2 - m^2}$. The spinors $|i\rangle, |i]$ go therefore into their massless counterparts, while the spinors $|n_i\rangle, |n_i]$ vanish.

If two masses are equal, say $m_i = m_j = m$ and the third mass is zero, i.e. $m_k = 0$, as is the case when two massive fermions interact with a massless boson, then one needs to introduce the so called x-factor [7], which can be obtained by contracting $p_j|k] = mx|k\rangle$ with $\langle \xi|$. Here leg $|k]$ has positive helicity, for leg $|k\rangle$ with negative helicity one contracts $p_j|k\rangle = m\tilde{x}|k]$ with $[\xi|$.

The x-factor is in the high energy limit using (11):

$$x = \frac{\langle \xi | p_j | k]}{m \langle \xi k \rangle} \approx \frac{\langle \xi j \rangle [j k][k j]}{m \langle \xi k \rangle [k j]} \approx \frac{-\langle \xi i \rangle [i k][k j]}{-m \langle \xi i \rangle [i j]} = \frac{[i k][k j]}{m [i j]} \tag{15}$$

The amplitude then becomes with (9)-(11) in the high energy limit:

$$x \langle i^J \ j^K \rangle \approx \frac{[i k][k j]}{m [i j]} \begin{pmatrix} \langle i j \rangle & \langle i n_j \rangle \\ \langle n_i j \rangle & \langle n_i n_j \rangle \end{pmatrix} \approx \begin{pmatrix} 0 & \frac{-[k j]^2}{[i j]} \\ \frac{[i k]^2}{[i j]} & 0 \end{pmatrix}$$

This example shows that amplitude calculations go faster without using the explicit spinors in (4) and employing only $|i\rangle$ and $|n_i\rangle$ together with (5).

## 4. Comments on recursion relations

In this section we comment on recursion relations, which in spinor helicity with massless particles are an important tool for calculating higher tree amplitudes, see [1-4] and [13]. We follow the discussion in [3] and [4]. In the soft limit of the propagator $P \to 0$ any amplitude can be factorized in smaller amplitudes. One deforms at least two momenta $p_i$ and $p_j$ by a complex variable z in a way, that momentum conservation and onshell conditions are guaranteed. This is the case if the following equations are satisfied:

$$\hat{p}_i = p_i - zq, \ \hat{p}_j = p_j + zq \tag{16}$$

$$q^2 = p_i \cdot q = p_j \cdot q = 0 \tag{17}$$

With the first equations (16) momentum conservation is satisfied due to $\hat{p}_i + \hat{p}_j = p_i + p_j$. With the next equations (17) the onshell condition is satisfied due to $\hat{p}_i^2 = (p_i - zq)^2 = p_i^2 - 2zp_i \cdot q + q^2 = p_i^2$ and similar for $p_j$. q must be a nullvector and orthogonal to $p_i$ and $p_j$. If both particles are massless one can choose $q = |i\rangle[j|$ satisfying the equations in (17). The momentum spinors then are shifted according to

$$|\hat{i}\rangle = |i\rangle, \ |\hat{i}] = |i] - z|j], \ |\hat{j}\rangle = |j\rangle + z|i\rangle, \ |\hat{j}] = |j] \tag{18}$$

One then has $\hat{p}_i = |\hat{i}\rangle[\hat{i}| = |i\rangle[i| - z|i\rangle[j|$ and $\hat{p}_j = |\hat{j}\rangle[\hat{j}| = |j\rangle[j| + z|i\rangle[j|$ realizing (16). This is the shift leading to the BCFW recursion [13]. The amplitude now becomes complex and can be calculated with the residue theorem, for details see [3] and [4]. The poles contributing to the residues are from keeping the propagator momentum $P_I(z) = P_I - zq$ onshell: $P_I(z)^2 = P_I^2 - 2zP_I \cdot q = M^2 \Rightarrow z = z_I = \frac{P_I^2 - M^2}{2q \cdot P_I}$. If the boundary contribution is zero, the amplitude can be written as $\mathcal{A}(0) = \sum_{z_I} \mathcal{A}_L(z_I) \frac{1}{P_I^2 - M^2} \mathcal{A}_R(z_I)$.

The conditions (16) and (17) are not easy to satisfy if one or two particles have mass, the simple generalization of (18) does not work as discussed in [11]. We first discuss the cases, when one particle is massive and the other massless.

**Case I:** $m_i = 0, m_j \neq 0$

The momenta are given as $p_i = |i\rangle[i|$ and $p_j = |j\rangle[j| + |n_j\rangle[n_j|$. From inspecting (18) it is clear, that one needs an $SU(2)$ vector $a^I$, to implement a shift of $|i]$ with $|j_I]$. We make an ansatz for the shifts analogue to (18)

$$|\hat{i}\rangle = |i\rangle, \ |\hat{i}] = |i] - za^I|j_I], \ |\hat{j}^I\rangle = |j^I\rangle + za^I|i\rangle, \ |\hat{j}_I] = |j_I] \tag{19}$$

One sees that momentum conservation is satisfied
$\hat{p}_i = |\hat{i}\rangle[\hat{i}| = |i\rangle[i| - za^I|i\rangle[j_I|$, $\hat{p}_j = |\hat{j}^I\rangle[\hat{j}_I| = |j^I\rangle[j_I| + za^I|i\rangle[j_I|$ and one obtains for the vector $q$, trivially satisfying $q^2 = 0 = q \cdot p_i$:

$$q = a^I |i\rangle [j_I| \tag{20}$$

From $q \cdot p_j = 0$ we get a condition for the vector $a^I$: $2q \cdot p_j = \langle j^J i\rangle a^I [j_I \, j_J] = \langle j^J i\rangle a^I \cdot -m_j \epsilon_{IJ} = +m_j \langle j_I \, i\rangle a^I = 0$. Using (3) and $a^I = (a^1 \; a^2)$ one obtains $\langle n_j \, i\rangle a^1 - \langle j\, i\rangle a^2 = 0$. In order to get as correct limit the BCFW recursion for $\langle n_j| = 0$ one puts $a^1 = 1, a^2 = \langle n_j \, i\rangle/\langle j\, i\rangle = a$.

$$a^I = (1 \; a) = \frac{\langle i \; j^I \rangle}{\langle i \; j \rangle}, \; a = \frac{\langle n_j \, i\rangle}{\langle j \, i\rangle}, \; a_I = (a \; -1) = \frac{\langle i \; j_I \rangle}{\langle i \; j \rangle}, \; a^I \langle j_I \, i\rangle = 0 \tag{21}$$

## Case II: $m_i \neq 0, m_j = 0$

This case is not entirely trivial, as one would think first, so we discuss it separately. The momenta are now given as $p_i = |i\rangle[i| + |n_i\rangle[n_i|$ and $p_j = |j\rangle[j|$. We make a similar ansatz for the shifts and use another vector $b_I$, which will turn out to be different from $a_I$.

$$|\hat{i}^I\rangle = |i^I\rangle, \; |\hat{i}_I] = |i_I] - zb_I|j], \; |\hat{j}\rangle = |j\rangle + zb_I|i^I\rangle, \; |\hat{j}] = |j] \tag{22}$$

First we check momentum conservation: $\hat{p}_i = |\hat{i}^I\rangle[\hat{i}_I| = |i^I\rangle[i_I| - zb_I|i^I\rangle[j|$, $\hat{p}_j = |\hat{j}\rangle[\hat{j}| = |j\rangle[j| + zb_I|i^I\rangle[j|$. Thereby one sees that vector $q$ is now defined as

$$q = b_I|i^I\rangle[j| = -b^I|i_I\rangle[j| \tag{23}$$

$q^2 = 0 = q \cdot p_j$ are trivially valid and from $q \cdot p_i = 0$ we can determine the vector $b_I$: $2q \cdot p_i = b_I \langle i^K \, i^I\rangle[j \, i_K] = b_I \cdot -m_i \epsilon^{KI}[j \, i_K] = +m_i b^K[j \, i_K] = 0$. This gives $b^1[j\,i] + b^2[j\,n_i] = 0$, now we put $b^2 = 1, b^1 = -[j\,n_i]/[j\,i] = b$ again in order to get the correct limit for $|n_i] = 0$. In summary we have

$$b^I = (b \; 1) = \frac{[j \, i^I]}{[j \, i]}, \; b = \frac{-[j \, n_i]}{[j \, i]}, \; b_I = (1 \; -b) = \frac{[j \, i_I]}{[j \, i]}, \; b^I[j \, i_I] = 0 \tag{24}$$

## Case III: $m_i \neq 0, m_j \neq 0$

Since we have obtained different vectors $a^I$ and $b^I$ in the two previous cases and we have seen the combinations $a^J|j_J]$ and $b_J|i^J]$ in (19) and (22) we try to retain them in the ansatz for two massive spinors:

$$|\hat{i}^I\rangle = |i^I\rangle, \; |\hat{i}_I] = |i_I] - zb_I a^J|j_J], \; |\hat{j}^I\rangle = |j^I\rangle + za^I b_J|i^J\rangle, \; |\hat{j}_I] = |j_I] \tag{25}$$

We see that momentum conservation is satisfied for the shifted momenta i.e. $\hat{p}_i + \hat{p}_j = p_i + p_j$:
$\hat{p}_i = |\hat{i}^I\rangle[\hat{i}_I| = |i^I\rangle[i_I| - zb_I a^J|i^I\rangle[j_J|$, $\hat{p}_j = |\hat{j}^I\rangle[\hat{j}_I| = |j^I\rangle[j_I| + za^I b_J|i^J\rangle[j_I|$
The vector $q$ is given by

$$q = b_I a^J|i^I\rangle[j_J| = -a^I b^J|i_J\rangle[j_I| \tag{26}$$

At first one has to check the asymptotic. For $m_i = 0, m_j \neq 0$ one gets for $|i_I] = (|i] \quad 0)$ and $|n_i] = 0$ from (25) by comparing with (19) $|\hat{i}_I] = (|\hat{i}] \quad 0) = (|i] \quad 0) - z(b_1 \quad b_2) a^J |j_J] \overset{!}{=} (|i] - za^J |j_J] \quad 0) \Rightarrow b_1 = 1, b_2 = 0$. The other shifts then automatically coincide with (19). The equation for $a^I$ is for compatibility with (18).

$$m_i = 0, m_j \neq 0 \Rightarrow n_i = 0, b^I = (0 \quad 1), b_I = (1 \quad 0), q = |i\rangle a^J [j_J|, a^I = (1 \quad a) \tag{27}$$

Similarly for $m_i \neq 0, m_j = 0$ one gets for $|j^I\rangle = (|j\rangle \quad 0)$ and $|n_j\rangle = 0$ from (25) by comparing with (22) $|\hat{j}^I\rangle = (|\hat{j}\rangle \quad 0) = (|j\rangle \quad 0) - z(a^1 \quad a^2) b_J |i^J\rangle \overset{!}{=} (|j\rangle - zb_J |i^J\rangle \quad 0) \Rightarrow a^1 = 1, a^2 = 0$. Again the other shifts are identical with (22). The equation for $b^I$ makes the limit compatible with BCFW in (18).

$$m_i \neq 0, m_j = 0 \Rightarrow n_j = 0, a^I = (1 \quad 0), a_I = (0 \quad -1), q = b_I |i^I\rangle [j|, b^I = (b \quad 1) \tag{28}$$

Now we have to check the onshell conditions in (17). The first one $2q^2 = \text{Tr}\{q \cdot \bar{q}\} = 0$ with $\bar{q} = -a^K b^L |j_K][i_L|$ gives $2q^2 = a^I b^J a^K b^L \langle i_L i_J\rangle [j_J j_K] = a^I b^J a^K b^L \cdot m_i \epsilon_{LJ} \cdot -m_j \epsilon_{JK} = -m_i m_j a_K a^K b^J b_J = 0$ due to (8). The next one results in: $2q \cdot p_i = b_I a^J \langle i^K i^I\rangle [j_J i_K] = b_I a^J \cdot -m_i \epsilon^{KI} [j_J i_K] = m_i b^K a^J [j_J i_K] = 0$ and for the third one we get: $2q \cdot p_j = b_I a^J \langle j^K i^I\rangle [j_J j_K] = b_I a^J \cdot -m_j \epsilon_{JK} \langle j^K i^I\rangle = -m_j a^J b^J \langle j_I i_J\rangle = 0$. So in summary we get two onshell conditions, which should determine the 2-vectors $a^I, b^I$

$$a^I b^J [j_I i_J] = 0 \tag{29}$$

$$a^I b^J \langle j_I i_J\rangle = 0 \tag{30}$$

Inspecting equations (21) and (24) we make the ansatz

$$a^I = (1 \quad a), b^I = (b \quad 1) \tag{31}$$

Inserting this in (29), (30) using (3) and evaluating the dot products between $a^I, b^J$ and the spinors $j_I, i_J$ one obtains two equations for a and b:

$$b[j i] + [j n_i] + ab[n_j i] + a[n_j n_i] = 0 \tag{32}$$

$$b\langle n_j n_i\rangle - \langle n_j i\rangle - ab\langle j n_i\rangle + a\langle j i\rangle = 0 \tag{33}$$

We again check the two limiting cases. For the case described in (27) with $b = 0$ and $n_i = 0$ one obtains from (32) $0 = 0$ and from (33) $a = \langle n_j i\rangle / \langle j i\rangle$. For the case in (28) with $a = 0$ and $n_j = 0$ one gets from (32) $b = -[j n_i]/[j i]$ and from (33) $0 = 0$, therefore these equations contain the limiting cases where one particle is massless and one massive. One can solve (32), (33) by solving both for a, and equate them and similarly for b, resulting in quadratic equations for a and b:

$$a^2 \langle j|p_i|n_j] + a(\langle j|p_i|j] - \langle n_j|p_i|n_j]) - \langle n_j|p_i|j] = 0 \tag{34}$$

$$b^2 \langle n_i|p_j|i] - b(\langle i|p_j|i] - \langle n_i|p_j|n_i]) - \langle i|p_j|n_i] = 0 \tag{35}$$

The solution of (34) under the condition $n_j \neq 0$ is:

$$a = \frac{-\langle j|p_i|j] + \langle n_j|p_i|n_j] + \sqrt{(\langle j|p_i|j] - \langle n_j|p_i|n_j])^2 + 4\langle j|p_i|n_j]\langle n_j|p_i|j]}}{2\langle j|p_i|n_j]} \quad (36)$$

In the case (27) $m_i = 0, m_j \neq 0, n_i = 0, b = 0, p_i = |i\rangle[i|$ only the plus sign in front of the square root gives the correct value $a = \langle n_j\ i\rangle/\langle j\ i\rangle$, if $\langle j|p_i|j] + \langle n_j|p_i|n_j] > 0$. The solution of (35) requiring $n_i \neq 0$ is

$$b = \frac{\langle i|p_j|i] - \langle n_i|p_j|n_i] - \sqrt{(\langle i|p_j|i] - \langle n_i|p_j|n_i])^2 + 4\langle i|p_j|n_i]\langle n_i|p_j|i]}}{2\langle n_i|p_j|i]} \quad (37)$$

In the case (28) $m_i \neq 0, m_j = 0, n_j = 0, a = 0, p_j = |j\rangle[j|$ only the minus sign gives the value $b = -[j\ n_i]/[j\ i]$.

One can write this in a more compact form. First one observes with the aid of (3), that the term before the square root in (36) can be written as $-\langle j^1|p_i|j_1] + 2\langle n_j|p_i|n_j]$, while the term under the square root in (36) can be written as $\langle j^1|p_i|j_1]^2 - 2\langle j^1|p_i|j^J]\langle j_1|p_i|j_J]$. With the relations in the third last and penultimate line of (7) one can write equation (36) as (if $p_ip_j > 0 \Rightarrow +\sqrt{\Delta}$, if $p_ip_j < 0 \Rightarrow -\sqrt{\Delta}$)

$$a = \left(\langle n_j|p_i|n_j] - p_ip_j + \sqrt{\Delta}\right)/\langle j|p_i|n_j], \text{ where } \Delta = (p_ip_j)^2 - m_i^2 m_j^2 \quad (38)$$

The limiting case (21) with $m_i = 0, p_i = |i\rangle[i|$ is now easily obtained as $a = \langle n_j|p_i|n_j]/\langle j|p_i|n_j] = \langle n_j\ i\rangle/\langle j\ i\rangle$. And similarly equation (37) can be written as (if $p_ip_j > 0 \Rightarrow -\sqrt{\Delta}$, if $p_ip_j < 0 \Rightarrow +\sqrt{\Delta}$)

$$b = \left(-\langle n_i|p_j|n_i] + p_ip_j - \sqrt{\Delta}\right)/\langle n_i|p_j|i] \quad (39)$$

The limiting case (24) with $m_j = 0, p_j = |j\rangle[j|$ is then obtained as $b = -\langle n_i|p_j|n_i]/\langle n_i|p_j|i] = -[j\ n_i]/[j\ i]$. Surprisingly (38), (39) bear some formal similarity with the approach in [14]. Finally we note, that now one can write the shift vectors $a^I, b^I$ in the more covariant form

$$a^I = (1\ \ a) = \left(\langle j^I|p_i|n_j] - \delta_2^I\left(p_ip_j - \text{sign}(p_ip_j)\sqrt{\Delta}\right)\right)/\langle j|p_i|n_j] \quad (40)$$

$$b^I = (b\ \ 1) = \left(\langle n_i|p_j|i^I] + \delta_1^I\left(p_ip_j - \text{sign}(p_ip_j)\sqrt{\Delta}\right)\right)/\langle n_i|p_j|i] \quad (41)$$

With the shifts in (25), the vector q in (26) and the solutions (36), (37) momentum conservation and onshell conditions are satisfied for particles with mass. In the case $m_i = 0, m_j \neq 0$ the shift vector $a^I = \langle i\ j^I\rangle/\langle i\ j\rangle$ (up to the factor guaranteeing here the BCFW limit) is like the shift in [15], where it was applied to a $W\bar{W}\gamma\gamma$ amplitude. See also the recent paper [16] concerning the case $m_i \neq 0, m_j = 0$ applied to several amplitudes, where the factor $1/[i\ j]$ in (24) is replaced by $1/m_i$. Therefore in the cases I and II, with one massless and one massive spinor, the application to amplitudes seems doable. In the case III with two massive spinors this certainly becomes more difficult.

## 4. Summary

In summary we have considered massive spinors and formulated them as 2-vectors, which makes it easy to obtain a couple of relations between them. We avoid entirely the display of $SL(2,\mathbb{C})$ indices, which simplifies many formulas considerably. An example for expanding a three particle amplitude in the high energy limit is shown. Finally we comment on recursion relations for massive spinors and show that it is possible to maintain momentum conservation and onshell conditions. The application of these spinor shifts to concrete amplitudes however is left as problem yet to be solved.

# Appendix A

Here we provide another explicit representation of massive spinors based on the standard momentum

$$p^\mu = \begin{pmatrix} E & P\sin(\theta)\cos(\phi) & P\sin(\theta)\sin(\phi) & P\cos(\theta) \end{pmatrix} \tag{A1}$$

Using the Pauli matrices, the momentum can be written in spinor notation $p = p_{\alpha\dot\alpha} = p_\mu \sigma^\mu$ and $\bar p = p^{\dot\alpha\alpha} = p_\mu \bar\sigma^\mu$,

$$p = p_{\alpha\dot\alpha} = \begin{pmatrix} E - P(cc-ss^*) & -2Pcs^* \\ -2Pcs & E + P(cc-ss^*) \end{pmatrix}, \quad \bar p = p^{\dot\alpha\alpha} = \begin{pmatrix} E + P(cc-ss^*) & 2Pcs^* \\ 2Pcs & E - P(cc-ss^*) \end{pmatrix} \tag{A2}$$

In analogy to (3) we now write the massive spinors in the 2-vector notation

$$|i^1\rangle = (|i\rangle \quad |n_i\rangle)\,, \quad \langle i^1| = (\langle i| \quad \langle n_i|)\,, \quad [i^1| = (-|n_i] \quad |i])\,, \quad [i^1| = (-[n_i| \quad [i|) \tag{A3}$$

$$|i_I\rangle = (|n_i\rangle \quad -|i\rangle)\,, \quad \langle i_I| = (\langle n_i| \quad -\langle i|)\,, \quad |i_I] = (|i] \quad |n_i])\,, \quad [i_I| = ([i| \quad [n_i|)$$

The spinors i and $n_i$ are now explicitly given as

$$|i\rangle = \sqrt{E_i + P_i}\begin{pmatrix} -s_i^* \\ c_i \end{pmatrix}\,, \quad |n_i\rangle = \sqrt{E_i - P_i}\begin{pmatrix} -c_i \\ -s_i \end{pmatrix}\,, \quad \langle i| = \sqrt{E_i + P_i}\begin{pmatrix} c_i \\ s_i^* \end{pmatrix}\,, \quad \langle n_i| = \sqrt{E_i - P_i}\begin{pmatrix} -s_i \\ c_i \end{pmatrix} \tag{A4}$$

$$|i] = \sqrt{E_i + P_i}\begin{pmatrix} c_i \\ s_i \end{pmatrix}\,, \quad |n_i] = \sqrt{E_i - P_i}\begin{pmatrix} -s_i^* \\ c_i \end{pmatrix}\,, \quad [i| = \sqrt{E_i + P_i}\begin{pmatrix} -s_i \\ c_i \end{pmatrix}\,, \quad [n_i| = \sqrt{E_i - P_i}\begin{pmatrix} -c_i \\ -s_i^* \end{pmatrix}$$

The momentum is still $p_i = |i^1\rangle[i_I| = |i\rangle[i| + |n_i\rangle[n_i|$ or $\bar p_i = |i_I]\langle i^1|$ and the relations in (5) remain valid.

$$\langle p_i \, n_i\rangle = -m_i\,, \quad [p_i \, n_i] = +m_i \tag{A5}$$

# References

[1] H. Elvang, Y. Huang, Scattering Amplitudes in Gauge Theory and Gravity, Cambridge University Press 2015
[2] J. M. Henn, J. C. Plefka, Scattering Amplitudes in Gauge Theories, Springer 2014
[3] T. R. Taylor, arXiv: 1703.05670 [hep-th]
[4] C. Cheung, arXiv: 1708.03872 [hep-ph]
[5] B. R. Holstein, Am.J.Phys. 74 (2006) 1002, arXiv: gr-qc/0607045
[6] C. Schwinn, S. Weinzierl, JHEP 0505 (2005) 006, hep-th/0503015
[7] N. Arkani-Hamed, T. C. Huang, Y. t. Huang, arXiv: 1709.04891 [hep-th]
[8] A. Ochirov, JHEP 1804 (2018) 089, arXiv: 1802.06730 [hep-ph]
[9] H. Johannsson, A. Ochirov JHEP 1909 (2019) 040 arXiv: 1906.12292 [hep-th]
[10] N. Christensen, B. Field, Phys. Rev. D 98, 016014 (2018), arXiv: 1802.00448 [hep-ph]
[11] N. Christensen, B. Field, A. Moore, S. Pinto, Phys. Rev. D 101, 065019 (2020), arXiv: 1909.09164 [hep-ph]
[12] H. K. Dreiner, H. E. Haber, S. P. Martin, Phys. Rept. 494, 1 (2010), arXiv: 0812.1594
[13] R. Britto, F. Cachazo, B. Feng, E. Witten, Phys. Rev. Lett. 94, 181602 (2005), hep-th/0501052
[14] C. Schwinn, S. Weinzierl, JHEP 0704 (2007) 072, hep-ph/0703021
[15] R. Aoude, C. S. Machado, JHEP 12 (2019) 058, arXiv: 1905.11433 [hep-ph]
[16] S. Ballav, A. Manna, arXiv: 2010.14139 [hep-th]
[17] The present version 2 of this paper was posted to viXra about a week earlier